\documentclass[useAMS,usenatbib]{mn2e}

\usepackage[usenames,dvipsnames]{xcolor}

\usepackage{aas_macros}
\usepackage{amssymb,amsmath}
\usepackage{epsf}
\usepackage{bm}
\usepackage{natbib}
\usepackage{graphicx}
\usepackage{cancel}
\usepackage{ulem}

\newcommand{\rot}{\mathrm{rot}\,}
\newcommand{\delt}[1]{\delta^{(\mathrm #1)}}

\title[R-modes in magnetized neutron stars]
{Differential rotation and r-modes in magnetized neutron
stars }

\author[Andrey I. Chugunov]{Andrey I. Chugunov$^1$\thanks{andr.astro@mail.ioffe.ru}\\
$^1$ Ioffe Institute, Polytekhnicheskaya 26, 194021
St.-Petersburg, Russia}

\begin{document}

\date{Accepted 2014 xxxx. Received 2014 xxxx;
in original form 2014 xxxx}

\pagerange{\pageref{firstpage}--\pageref{lastpage}}
\pubyear{2014}

\maketitle

\label{firstpage}

\begin{abstract}
Rezzolla et al. [ApJ \textbf{531} (2000), L139; Phys.\
Rev.\ D \textbf{64} (2001), 104013; Phys.\ Rev.\ D
\textbf{64} (2001), 104014] draw attention to the second
order secular drift associated with r-modes and claimed
that it should lead to magnetic field enhancement and
suppression of r-mode instability in magnetized neutron
stars. We critically revise these results. We present a
particular second order r-mode solution with vanishing
secular drift, thus refuting a widely believed statement
that secular drift is an unavoidable feature of r-modes.
This non-drifting solution is not affected by magnetic
field $B$, if $B\ll B_{\mathrm{crit}}\approx
10^{17}\,(\nu/600\,\mathrm{Hz})$~G ($\nu$ is a spin
frequency) and does not lead to secular evolution of
magnetic field. For general second order r-mode solution
the drift does not necessarily  vanish, but the solution
can be presented as a superposition of two solutions: one
describes evolution of differential rotation in
nonoscillating star (which describes secular drift; for
nonmagnetized star it is arbitrary stationary rotation
stratified on cylinders; for magnetized star differential
rotation evolves on the Alfv\'{e}n timescale and may lead
to magnetic energy enhancement), and another one is
non-drifting r-mode solution mentioned above. This
representation allows us to conclude that enhancement of
magnetic field energy is limited by initial energy of
differential rotation, which is much less (for a factor
$\propto \alpha^2$, where $\alpha$ is mode amplitude) than
the total energy of r-mode. Hence, magnetic field
enhancement by drift cannot suppress r-mode instability.
Results can be generalized for any oscillation mode in any
medium, if this mode has non-drifting solution for $B=0$.
\end{abstract}

\begin{keywords}
stars: neutron -- stars: oscillations -- MHD -- dynamo
\end{keywords}

\section{Introduction}

Observations of millisecond pulsars demonstrate that some
neutron stars (NSs) rotate very rapidly (the fastest known
pulsar PSR J1748-2446ad has spin frequency
$\nu=\Omega/2\pi\approx 716$~Hz; \citealt{hrsfkc06}).
However, as was shown by \cite{andersson98} and
\cite{fm98}, rapidly rotating NSs are subject to
gravitation-driven instability associated with enhancement
of r-modes (toroidal oscillation mode of rotating star
controlled by Coriolis force). It is a particular case of
the Chandrasekhar-Friedman-Schutz (CFS) instability
[\citealt{chandrasekhar70a,fs78a,fs78b}]. R-mode
instability probably plays an important role in NS physics.
Indeed, it is argued to be a mechanism which limits
pulsars' spin frequencies (\citealt*{bildsten98,aks99,gck14a,gck14b}). Rapidly rotating hot NSs, observed
in low mass X-ray binaries (LMXBs), can be affected by
r-mode instability even more profoundly: some of these
stars should be unstable within the standard model of
r-mode oscillations (see \citealt*{hah11,hdh12};
\citealt{gck14a}, for example) and their existence provides
one with a good opportunity to test microphysical models of
NS's core (see, e.g., \citealt{hah11,hdh12,gck14a,gck14b}).
Furthermore, \cite*{cgk14} suggest that r-mode instability
can support a new class of rapidly rotating NSs, which do
not accrete matter from companion even transiently (as NSs
in LMXBs), but still keep high temperature via r-mode
instability. Some of these NSs might have been already
observed, but erroneously classified as quiescent LMXB
candidates (see \citealt*{cgk14} for detailed discussion).
Finally, as argued by \cite{Rezzolla_etal00}, r-mode
instability can be also important for generation of NS
magnetic fields. The main aim of the paper is to discuss
the role, which magnetic field plays in nonlinear evolution
of r-modes and role of r-modes in magnetic field evolution
in NSs.

Most of the papers discussing r-modes in NSs either neglect
magnetic field or simply mention that it can be important.
Indeed, dipolar magnetic field of rapidly rotating NSs
($\nu\gtrsim 200$~Hz) is not very large $\sim 10^8$~G and
seems to be clearly negligible (see
\citealt*{Rezzolla_etal00,aly15} and Sec.\ \ref{Sec:mag}).
Recent papers (\citealt*{lee05,arr12,cs13,aly15}, for
instance) confirm by detailed numerical calculations that
only very high magnetic field can affect r-modes strongly
[for example (see fig. 7 by \cite{cs13}), magnetic field
$B\gtrsim 3\times 10^{15}$~G modifies r-mode frequency for
a less than for a one percent for $\nu\approx 220$~Hz].
However, just after r-mode instability was discovered,
\cite{Rezzolla_etal00,Rezzolla_etal01a,Rezzolla_etal01b}
pointed that r-modes are able to increase seeding magnetic
field through secular drift of fluid elements, which takes
place within second order (in oscillation amplitude)
perturbation theory. They argue that this mechanism can be
responsible for generation of strong magnetic fields in
NSs, but at the same time it finally suppresses r-mode
instability. A strong support for these results was
provided by subsequent papers, which claim differential
rotation, i.e. secular drift of fluid elements, to be a
necessary feature of r-modes (in nonmagnetized NS) on the
base of analytical (\citealt{Sa04,ST05}) and numerical
calculations (see \citealt{SF01,LTV01}, for example).

In this paper we reanalyze these results using analytical
derivations within second order (in oscillation amplitude)
perturbation theory.  In Sec.\ \ref{Sec:nonmag} we discuss
nonmagnetized ($B=0$) NSs, following the approach suggested
by \cite{Sa04,ST05}. As they have shown, the second order
r-mode solution is not unique, but determined up to any
differential rotation, which can exist in nonoscillating
star (i.e. arbitrary stationary  differential rotation,
stratified on the cylinders). However, an important point
of these papers, namely that ``differential rotation,
producing large scale drifts of fluid elements along
stellar latitudes, is an unavoidable feature of r-modes in
the nonlinear theory'' (see abstract in \citealt{ST05}) is
not true: we present a parameter set for analytical second
order solution obtained by \cite{Sa04,ST05}
with vanishing secular drift (below we refer to it as to ``non-drifting solution'').%
\footnote{\cite{Sa04} and \cite{ST05} base their conclusion
on the oscillation averaged second order perturbations of
\textit{Eulerian} velocity. Indeed, these perturbations are
nonvanishing for all possible second order solutions.
However, the drift of fluid elements, i.e. secular increase
of the Lagrangian displacement, is not determined by the
second order Eulerian velocity only, but has correction
associated with Stokes drift induced by the first order
motion of the fluid elements (see
\citealt{LH53,Rezzolla_etal01a} or Eqs. (28) and (12) in
\citealt{Sa04} and \citealt{ST05}, respectively). This
correction allows to have second order solution with
vanishing secular drift, but nonvanishing oscillation
averaged Eulerian velocity (see Sec.\ \ref{Sec:nonmag} for
details).}
We argue that existence of second order solution with
vanishing drift is not coincidental feature, specific for
r-modes only, but rather is a general property of
oscillation modes. In Sec.\ \ref{Sec:mag} we reconsider the
effect of magnetic field on r-modes in second order in
amplitude by accurate treatment of magnetic stresses within
ideal magneto-hydrodynamics (MHD).
\cite{Rezzolla_etal00,Rezzolla_etal01a,Rezzolla_etal01b}
argued that r-mode velocity profile is not affected by
magnetic field up to $B\ll 10^{16}$~G.
Assuming secular drift to be unaffected, they conclude that
r-modes enhance magnetic field due to differential
rotation, associated with drift.%
\footnote{\label{footnote_af63} As it is stressed by
\cite{af63} (Sec.\ 3.14.5), the twist of magnetic field
caused by inhomogeneous rotation, propagates along magnetic
lines of force with Alfv\'{e}n velocity $v_\mathrm{A}$,
thus if rotation inhomogeneity is too small, the force
lines will have enough time to straighten up. As a result,
to produce significant field by differential rotation,
relative velocity of different parts of the star should be
at least comparable with Alfv\'{e}n velocity. However, this
point was not discussed by
\cite{Rezzolla_etal00,Rezzolla_etal01a,Rezzolla_etal01b}.
On the contrary, they suppose that arbitrary small secular
drift can enhance magnetic field.} %
We demonstrate that non-drifting solution is indeed
unaffected by magnetic field up to $B\ll
B_{\mathrm{crit}}\approx
10^{17}\,(\nu/600\,\mathrm{Hz})$~G, but it does not lead to
secular evolution of magnetic field. However, in a general
second order solution the second order velocity profile
stays unaffected by magnetic field only for a timescale
much less then the Alfv\'{e}n timescale (while
back-reaction of magnetic field can be neglected). On the
contrary, at the Alfv\'{e}n timescale the drift
contribution to the second order velocity profile
(corresponding to arbitrary stationary  differential
rotation) is modified by magnetic field [back-reaction is
crucial and cannot be neglected for the same reasons as why
stationary differential rotation is forbidden in magnetized
stars (Ferraro's law of isorotation)].
To describe back-reaction properly, we demonstrate that
general second order r-mode solution in a magnetized NS
with $B\ll B_{\mathrm{crit}}$ can be presented as a
superposition of two independent solutions of MHD
equations: (a) solution which describes evolution of
differential rotation in nonoscillating magnetized NS
(\textit{drift solution}, which describes evolution of
secular drift), and (b) non-drifting r-mode solution. These
solutions are decoupled (at second order in oscillation
amplitude) as far as Eulerian perturbations of all
quantities (except magnetic field)
in drift solution are second order in amplitude of
non-drifting r-mode solution.
Evolution of the drift solution is governed by the same
equations as
in \textit{non-oscillating} magnetized NS, thus it evolves
at Alfv\'{e}n timescale (see \citealt{Spruit99_DifRot}, for
example). Secular evolution of the magnetic field is solely
determined by the drift, thus initial energy of
differential rotation provides an upper bound for the
increase of magnetic field energy. Furthermore, energy of
the non-drifting mode is conserved in the leading order. As
a result, r-modes cannot convert their energy to magnetic
field. Hence we conclude, that magnetic energy enhancement
by secular drift cannot suppress r-mode instability.%
\footnote{As argued by \cite{mendell01,km03}, magnetic
field can affect r-mode damping at the crust-core boundary,
but it requires rather strong radial field ($> 10^{11}$~G),
which is much larger than typical dipolar magnetic fields
of rapidly rotating NSs ($\approx 10^8$~G). It is also
unlikely that strong \textit{radial} magnetic fields can be
generated by drift.}

\section{R-modes in nonmagnetized NS}
\label{Sec:nonmag}

In this section we consider r-modes oscillations in slowly
rotating Newtonian NSs with barotropic equation of state
(i.e. pressure depends only on density). In case of general
relativity oscillation equations become much more
complicated (see \citealt*{rk02,lfa03}, for example).
Furthermore, \cite*{laf00} demonstrate fully relativistic
equations for r-modes in NSs with barotropic and
nonbarotropic equations of state to be qualitatively
different. For barotropic equation of state a relativistic
analogue for Newtonian r-modes was found by \cite{lfa03}
and they conclude that ``unstable r-modes remain
essentially unaltered when the problem is studied in full
general relativity''.\footnote{
\cite{cs13} stated that their relativistic Cowling
calculations give r-mode gravitational radiation growth
time 30\% larger than in previous Newtonian results
(\citealt{lom98,aks99}). This statement seems to be
partially biased by different radius and mass of NS taken
as fiducial by \cite{cs13} ($R=14.08$~km and
$M=1.4M_\odot$) and by \cite{aks99} ($R=12.47$~km and
$M=1.5M_\odot$). Being rescaled to the fiducial parameters
of \cite{cs13} by applying scaling relations from
\cite{ak01} [$\tau_\mathrm{gr}\propto 1/(M R^4)$], results
by \cite{lom98,aks99} come to a much better agreement with
results by \cite{cs13}. %
Correspondent growth time is $\tau_\mathrm{gw}\approx 13.65
(1\mathrm{kHz}/\nu)^{5.83}$, $\tau_\mathrm{gw}\approx 12.47
(1\mathrm{kHz}/\nu)^6$, $\tau_\mathrm{gw}\approx 13.73
(1\mathrm{kHz}/\nu)^{5.93}$, for Refs.\ \cite{cs13},
\cite{lom98}, and \cite{aks99} respectively. ].} It allows
us to take advantage of Newtonian r-modes for the sake of
simplicity.

Second order (in oscillation amplitude) r-mode solution was
obtained by \cite{Sa04}.
This solution is applicable
if oscillation amplitude $\alpha\gg
(\Omega/\Omega_\mathrm{K})^2$ (here
$\Omega_\mathrm{K}\approx \sqrt{GM/R^3}$ is the mass
shedding frequency).
Several important features of  r-mode solution by
\cite{Sa04} can be generalized for rather general
oscillation mode (see below). Thus, below we do not
restrict our consideration to r-modes, but we apply Sa's
r-mode solution as an example.

Let's look at the equations describing oscillations up to
second order in oscillation amplitude $\alpha$. The first
order equations are:
\begin{eqnarray}
\partial_t \delt{1} v_i+\delt{1} v^k\nabla_k v_i+v^k\nabla_k
\delt{1}
v_i&=&-\nabla_i \delt{1} U, \label{lin_nonmag1}\\
\partial_t \delt{1} \rho+v^i\nabla_i \delt{1}
\rho+\nabla_i (\rho \delt{1} v^i)&=&0,\label{lin_nonmag2}\\
\triangle \delt{1} \Phi&=&4\pi G \delt{1} \rho,
\label{lin_nonmag3}
\end{eqnarray}
where $\bm v$, $\rho$, $\Phi$, $p$ are, respectively,
velocity, density, gravitational potential, and pressure in
the unperturbed NS. $\delt{i} f=\mathcal{O}(\alpha^i)$
represents the $i^\mathrm{th}$ order Eulerian perturbation
of quantity $f$, and $\delt{i}
U=\delt{i}p/\rho+\delt{i}\Phi$. Finally, $G$ is
gravitational constant. In leading order in $\Omega$,
linear r-mode solution  is well known (see
\citealt*{pbr81}, for example) and velocity
perturbation can be written in form%
\footnote{We use normalization of the mode amplitude from
\cite{olcsva98}, which is usually applied for r-mode
instability analysis. \cite{Sa04,ST05} apply slightly
different normalization, namely their amplitude
$\alpha_\mathrm{Sa}=\alpha/\sqrt{2}$.
}.
\begin{equation}
  \delt{1} {\bm v} =\alpha \frac{\Omega R r}{\sqrt{l (l+1)}} \left(\frac{r}{R}\right)^l
         { \nabla} \times(r { \nabla} Y_{lm}) \mathrm{e}^{\imath \omega t},
         \label{deltaVprofile}
\end{equation}
where $Y_{lm}$ is the spherical harmonic with the
multipolarity $l$ equal to $m$,
$Y_{lm}=Y_{ll}=%
\sqrt{(2l+1)!/[2^{l+2}\,\pi\, (l!)^2]}\,\sin^l(\theta)
\mathrm e^{\imath l\phi}$. Here $\omega$ is the oscillation
frequency in the inertial frame, given by (also to leading
order in $\Omega$)
\begin{equation}
\omega = - \frac{(l-1) (l+2)}{l+1} \, \Omega. \label{o}
\end{equation}

The second order (in $\alpha$) equations  are:
\begin{eqnarray}
\partial_t \delt{2} v_i&+&\delt{2} v^k\nabla_k v_i+v^k\nabla_k
 \delt{2} v_i+\delt{1} v^k \nabla_k \delt{1} v_i
  \nonumber\\
 &=&
 -\nabla_i \delt{2} U+\frac{\delt{1} \rho}{\rho}\nabla_i \left(\frac{\delt{1} p}{p}\right),
 \label{sec_nonmag1}\\
\partial_t \delt{2} \rho&+&v^i\nabla_i \delt{2}
\rho+\nabla_i (\rho \delt{2} v^i)
 \nonumber\\
&+& \nabla_i (\delt{1}\rho \delt{1} v^i)=0,
\label{sec_nonmag2} \\
\triangle \delt{2} \Phi&=&4\pi G \delt{2} \rho.
\label{sec_nonmag3}
\end{eqnarray}
Let us stress two features of these equations:
\begin{itemize}
\item For a given first order solution, general solution of second order equations is a sum
of partial solution of inhomogeneous equations [Eqs.
(\ref{sec_nonmag1}-\ref{sec_nonmag3}), which contains terms
induced by the first order solution] and a general solution
of homogeneous part of Eqs.
(\ref{sec_nonmag1}-\ref{sec_nonmag3}) [without terms
induced by the first order solution].

\item Homogeneous part of equations (\ref{sec_nonmag1}-\ref{sec_nonmag3}) have the same form as
linearized equations [Eqs.
(\ref{lin_nonmag1}-\ref{lin_nonmag3})]. Thus, evolution of
all quantities in arbitrary solution of homogeneous
equations is the same as for  (linear) perturbations of
initial (nonoscillating) state.
\end{itemize}
Thus, with accuracy up to the second order in $\alpha$, one
can select an arbitrary partial solution of the oscillation
equations and present a general solution as composition of
selected oscillating solution and solution, which describes
arbitrary initial $\mathcal{O}(\alpha^2)$ perturbations,
evolving in a same way as in nonoscillating NS. In fact, it
is obvious even without explicit equations, since second
order perturbations can neither affect oscillation solution
nor be affected by it (within $\alpha^2$ accuracy). Let's
restrict subsequent consideration to nonoscillating
solutions of homogeneous Eqs.\
(\ref{sec_nonmag1}-\ref{sec_nonmag3}). In case of r-modes
it is stationary differential rotation, stratified on
cylinders.
The freedom in a homogeneous part of the solution can be
also understood as excitation of r-mode in a differentially
rotating star. As far as this background differential
rotation is weak (second order in oscillation amplitude) it
does not affect r-mode. This result is in agreement with
\cite*{kye01}, who discuss r-mode oscillation of
differentially rotating polytropic stars and obtain r-mode
solution in a case, when degree of differential rotation is
not too high.
Recently, \cite*{csy13,csy14} confirm that r-modes are
rather insensitive to small differential rotation within
general relativity Cowling approximation. Note, however,
that in case of very strong differential rotation,
r-modes can be suppressed (see \cite{kye01}, for more
details).

Partial solution, mentioned above, can have drift, i.e.\
the Lagrangian displacements $\bm \xi$ can increase with
time producing large scale secular motion of fluid
elements. Corresponding velocity (so-called mass transfer
velocity; see \citealt{LH53}, for example) $\bm
v^{(\mathrm{mt})}=\overline{\partial_t \bm \xi}$ is second
order in oscillation amplitude. Here overline means
averaging over oscillations. Generally, mass transfer
velocity is not equal to oscillation averaged Eulerian
velocity, but has a correction associated with Stokes drift
velocity $\bm v^{(\mathrm S)}$ induced by first order pure
oscillating solution $\bm v^{(\mathrm S)}=\overline{(\bm
\xi^{(1)} \nabla)\delt{1} \bm v}$ (see \citealt{LH53}, for
details).
As a result, Eulerian velocity perturbations can be written
in form:
\begin{eqnarray}
    \delt 1v^i&=& \partial_t \xi^{(1)\,i}+v^j\nabla_j
    \xi^{(1)\,i}-\xi^{(1)\,j}\nabla_j v^i
  \nonumber \\
    &=&(\partial_t+\Omega\partial_\phi) \xi^{(1)\,i}, \label{delt1v}
  \\
    \delt 2 v^i&=& \partial_t \xi^{(2)\,i}+v^j\nabla_j
    \xi^{(2)\,i}-\xi^{(2)\,j}\nabla_j v^i
    \nonumber\\
    &-&\xi^{(1)\,j}\nabla_j \delt 1v^i
    - \frac{1}{2} \xi^{(1)\,j}\xi^{(1)\,k}\nabla_j \nabla_k
    v^i
    \nonumber\\
    &=&(\partial_t+\Omega\partial_\phi) \xi^{(2)\,i}
    -\xi^{(1)\,j}\nabla_j \delt 1v^i.
 \label{delt2v}
\end{eqnarray}
Here the last equality corresponds to solid state rotation
at frequency $\Omega$ for unperturbed state ($v^i=\Omega
\delta^i_\phi$). By averaging of these equations one can
obtain relations for mass transfer velocity.

For r-modes mass transfer velocity $\bm v^{(\mathrm{mt})}$
can be easily derived from Eqs. (30-31) in \cite{Sa04},
which gives the second order Lagrangian displacement. Only
$\phi$ component of $\bm v^{(\mathrm{mt})}$ is nonvanishing
and it can be written in form:\footnote{In Eqs. (30-31) by
\cite{Sa04} summation over $N$ is omitted, but it is
required  for general solution.}
\begin{equation}
  (v^{(\mathrm{mt})})^\phi=\alpha^2 \Omega\left[
  \tilde{A} \left(\frac{r\, \sin\theta}{R}\right)^{2l-2}
 +\sum_{N=0}^\infty A_N
 \left(\frac{r\, \sin\theta}{R}\right)^{N}
 \right],
 \label{vd_Sa}
\end{equation}
where $A_N$ are arbitrary constants, and
\begin{equation}
\tilde{A}=\frac{(2l+1)!}{2^{2l+4}\,\pi}\,\frac{(2l-1)\,l}{l!^2}.
\end{equation}
For $A_{2l-2}=-\tilde{A}$  and $A_N=0$ for all other $N$ we
obtain a second order r-mode solution with vanishing mass
transfer velocity.%
\footnote{Note, if $A_0\ne0$, it corresponds to uniform
rotation, i.e. perturbation of the rotation frequency of
the star. Such drift will not perturb magnetic field
energy. All results obtained for nondrifting solution can
be also applied for solutions, which differ from nondrifting
solution by second order uniform rotation.}
For this solution Lagrangian displacements do not increase
with time, thus secular drift of fluid elements vanishes.
However, oscillation averaged Eulerian velocities are not
zero for this solution (see Eqs. (18) and (21) in
\citealt{Sa04}). Such a solution corresponds to a very
specific set of initial conditions and it is not
surprising, that it was not obtained in numerical
calculations by \cite{SF01,LTV01}. Nonetheless, it
simplifies discussion of r-modes in magnetized NSs (see
Sec.\ \ref{Sec:mag}).

Presence of a second order solution with vanishing drift is
not an accidental specific feature of r-modes. This
property can be generalized for any oscillation mode, if
mass transfer velocity $\bm v^{(\mathrm{mt})}$ for initial
partial solution of second order Eulerian equations
corresponds to a solution of linear perturbation equations
(or, equally, homogeneous second order equations).%
\footnote{\label{footnote_gen} It is likely to be rather
general case. For example, for ocean waves (incompressible
fluid in external gravity field) this property was
demonstrated by \cite{LH53,Moore70}.
In particular, for r-modes this generalization may allow to
go beyond restriction of second order solution by
\cite{Sa04} ($\Omega\ll\Omega_\mathrm{K}$ and  $\alpha\gg
(\Omega/\Omega_\mathrm{K})^2$).
 However, we leave formal
discussion, whether it is a general property or not, beyond
the scope of the paper.
}
In this case one can construct a partial solution with
vanishing drift by substituting $\delta \bm v \rightarrow
\delta \bm v-\delta^{(\mathrm {d})}\bm v$, $\delta \rho
\rightarrow \delta \rho - \delta^{(\mathrm {d})} \rho$,
$\delta \Phi \rightarrow \delta \Phi - \delta^{(\mathrm d)}
\Phi$. Here $\delta^{(\mathrm d)} \rho$ and
$\delta^{(\mathrm d)} \Phi$ are perturbations of density
and gravitational potential in a solution of linear
perturbation equations, which corresponds to velocity
perturbation $\delta^{(\mathrm{mt})} \bm v$.

Concluding, in this section we demonstrate, that general
second order r-mode  solution can be presented as a sum of
oscillating solution with \textit{vanishing drift} and a
solution, which describes differential rotation in
\textit{nonoscillating} NS. As far as we consider
nonmagnetized star, this differential rotation can be
described by arbitrary stationary profile, stratified on
cylinders.
If we restrict our consideration to r-modes excited by the
gravitational radiation in initially uniformly rotated NSs,
this profile is well defined (\citealt*{fll15}).
However, stationary differential rotation is forbidden for
magnetized star (Ferraro isorotation law), thus for
magnetized NS a general r-mode solution cannot be presented
as sum of oscillations and free \textit{stationary}
differential rotation. Magnetic field modifies  general
r-mode solution and we discuss this modification in the
next section.

\section{R-modes in magnetized NS} \label{Sec:mag}

Oscillations in nonmagnetized NS can have nonvanishing
drift, leading to large (and increasing) Lagrangian
displacements, but Eulerian perturbation of all quantities
stays small. This is not the case for  a magnetized NS.
Magnetic field is frozen into fluid elements, so that large
Lagrangian displacements generally lead to large Eulerian
perturbations of magnetic field. As a result, one generally
cannot expand Eulerian perturbation of magnetic field in
powers on oscillation amplitude and apply perturbed
Eulerian approach to describe r-modes with nonvanishing
drift. Hence, a stricter formalism is
required to describe oscillations with nonvanishing secular
drift in
magnetized NS.%
\footnote{A pure Alfv\'{e}n waves are  linear at arbitrary
amplitude and one can use Eulerian perturbations of
magnetic field to describe these waves even for large
amplitudes. Still, it is not the case of r-modes.}
Here we describe such formalism based on the idea to
present a general second order r-mode solution as a
superposition of two solutions: (a) solution which
describes evolution of the differential rotation in
nonoscillating NS within exact MHD equations (later
referred to as 'drift solution', because it will describe
drift), and (b) non-drifting oscillating solution for
$B=0$, described in the previous section. To achieve that,
we apply perturbations in two steps. First, we apply
initial perturbation, which is second order in $\alpha$,
and evolve it using exact MHD equations. This solution
describes evolution of drift motion [i.e.\ it is the drift
solution mentioned above]. Second, we perturb drift
solution,
write down perturbation equations up to second order in
$\alpha$, and show that non-drifting oscillating solution
of amplitude $\alpha$ is a (partial) solution of these
equations.
Finally, we demonstrate that r-mode with arbitrary second
order differential rotation at initial moment can be
presented in form of this superposition, thus the
superposition can be applied to describe a general solution
in magnetized case. It is worth to note that this procedure
is suitable for any oscillation mode in any media, if the
mode have non-drifting solution for $B=0$.

Let $(\rho, p,\ \bm v,\ \bm B)$ be a stationary
configuration described by solution of ideal MHD equations:%
\footnote{Applicability of ideal MHD for NS r-modes is
justified by high conductivity of NS depths (see
\citealt{ys91}, for example).}
\begin{eqnarray}
\partial_t \bm v+(\bm v\nabla) \bm v&=&
-\frac{\nabla p}{\rho}-\nabla \Phi +\frac{1}{4\pi\rho} [\bm
B\times [\nabla\times \bm B]],
 \label{MHD1}\\
\partial_t \rho+\nabla (\rho \bm v)&=&0, \label{MHD2}\\
\triangle  \Phi&=&4\pi G  \rho,  \label{MHD3}\\
 \frac{\partial B}{\partial t}&=&
 [\nabla\times[\bm v\times \bm B]].
\end{eqnarray}
Let $\bm c(\bm a, t)$ be a fluid element trajectory,
corresponding to this stationary solution.%
\footnote{Strictly speaking, we require NS to be static in
a rotating frame (corotating with star). We do not require
magnetic field to be axially symmetric with the rotation
axis.
We neglect deformations of a star due to magnetic stresses.
\cite{aly15} show these deformations to be rather
unimportant for r-modes.}
Solution of the induction equation can be written in the
form (see, e.g., \citealt{Rezzolla_etal01a})
\begin{equation}
\frac{B^j(\bm c(\bm a, t),t)}{\rho(\bm c(\bm a,
t),t)}=\frac{B^i(\bm a,t=0)}{\rho(\bm a,t=0)}\frac{\partial
\bm c^j(\bm a,t)}{\partial a^i}. \label{ind}
\end{equation}
For the sake of simplicity we assume $c^i(\bm a,t)= a^i+\bm
v^i t$, thus, stationary condition implies $\rho(\bm c(\bm
a, t),t)=\rho(\bm a, t)$,
i.e. the density $\rho$
is constant along trajectory $\bm c(\bm a, t)$.

As the first step let us consider the drift solution. We
define it as a perturbation of unperturbed stationary
configuration which is an \textit{exact} solution of MHD
equations. Let drift solution be given by the trajectory
$\bm c^{(\mathrm d)}(t)=\bm c(\bm a, t)+\bm \xi^{(\mathrm
d)}(\bm a, t)$ and Eulerian variables
  $(\rho^{(\mathrm d)}(\bm r, t),\
     p^{(\mathrm d)}(\bm r,\, t),\
     \bm v^{(\mathrm d)}(\bm r,t),\
     \bm B^{(\mathrm d)}(\bm r,\,t))
  $.
However, let us assume that Eulerian perturbations of
$\Phi$, $\rho$, $p$, $\bm v$ with respect to unperturbed
stationary configuration, $\delt{d} f=f^{(\mathrm d)}-f$,
are second order in $\alpha$:
$\delt{d}f=\mathcal{O}(\alpha^2)$ (at least for a time of
consideration) and their space and time derivatives have
same order in $\alpha$ (i.e. they are not too large:
$\nabla_i \delt{d} f\lesssim \delt{d} f/R$; $\partial_t
\delt{d} f\lesssim \Omega
\delt{d} f$).%
Here $f$ is $\Phi$, $\rho$, $p$, $\bm v$ and $\alpha\ll 1$.
We do not assume Eulerian perturbations of magnetic field
to be small and thus do not expand them in series on
$\alpha$. Instead, we consider evolution of total magnetic
field, which is given by the equation:
\begin{eqnarray}
  \frac{(\bm B^\mathrm{(\mathrm d)})^j(\bm r,t)}{\rho^\mathrm{(\mathrm d)}(\bm r, t)}
  &=&\frac{B^i(\bm a,t=0)}{\rho(\bm a,t=0)}
  \frac{\partial (\bm c^{(\mathrm{d})})^j(\bm a, t)}{\partial  a^i}
  \nonumber \\
 &=&\frac{B^i(\bm a,t=0)}{\rho(\bm a,t=0)}
  \left[\delta^j_i+\frac{\partial (\bm \xi^{(\mathrm d)})^j(\bm a,t)}{\partial\bm  a^i}\right]. \label{Ind_Bd}
\end{eqnarray}
Let us now perturb \textit{drift} solution by rapidly
oscillating perturbation of amplitude $\alpha$. To describe
its evolution, we apply perturbed (with respect to the
drift solution) MHD equations of the first and the second
order in Eulerian form.
In these equations one can substitute $\rho^{(\mathrm d)}$,
$p^{(\mathrm d)}$, $\Phi^{(\mathrm d)}$, and $\bm
v^{(\mathrm d)}$ with correspondent values of unperturbed
\textit{stationary} configuration because they differ only
in the second order
in $\alpha$.
Thus, linear perturbation equations can be written in form:
\begin{eqnarray}
\partial_t \delt{1} v_i+v^k\nabla_k \delt{1} v_i+v^k\nabla_k
\delt{1} v_i&=& -\nabla_i \delt{1} U \nonumber +\delt{1}
F^\mathrm{m}_i,
\nonumber \\ \label{lin_mag1}\\
\partial_t \delt{1} \rho+v^i\nabla_i \delt{1}
\rho+\nabla_i (\rho \delt{1} v^i)&=&0,  \label{lin_mag2}\\
\triangle \delt{1} \Phi&=&4\pi G \delt{1} \rho.
\label{lin_mag3}
\end{eqnarray}
Magnetic stress of the first order is:
\begin{eqnarray}
 \delt{1} \bm F^\mathrm{m}&=&
 \frac{1}{4\pi\rho}\left\{
   [\delt{1}\bm B\times\rot \bm B^{(\mathrm d)}]
   +[\bm B^{(\mathrm d)}\times \rot \delt{1} \bm B]
   \right\} \nonumber \\
 &-&\frac{\delt{1}\rho}{4\pi \rho^2}[\bm B^{(\mathrm d)}\times\rot \bm B^{(\mathrm d)}].
 \label{MagStress1}
\end{eqnarray}
The second order equations in $\alpha$ are:
\begin{eqnarray}
\partial_t \delt{2} v&+&v^k\nabla_k \delt{2} v_i+v^k\nabla_k
 \delt{2} v_i+\delt{1} v^k \nabla_k \delt{1} v_i \nonumber
 \\&=&
 -\nabla_i \delt{2} U+\frac{\delt{1} \rho}{\rho}\nabla_i \left(\frac{\delt{1} p}{p}\right)
 +\delt{2} F^\mathrm{m}_i,  \label{sec_mag1}\\
\partial_t \delt{2} \rho&+&v^i\nabla_i \delt{2}
\rho+\nabla_i (\rho \delt{2} v^i)=0,  \label{sec_mag2}\\
\triangle \delt{2} \Phi&=&4\pi G \delt{2} \rho,
\label{sec_mag3}
\end{eqnarray}
where second order magnetic stress is:
\begin{eqnarray}
\delta^{2} \bm F^\mathrm{m}&=&\frac{1}{4\pi
\rho}\left\{[\delt{2}\bm B\times\rot \bm B^{(\mathrm
d)}]+[\bm B^{(\mathrm d)}\times \rot \delt{2} \bm
B]\right\}
\nonumber\\
& +&
 \frac{1}{4\pi \rho}[\delt{1}\bm B\times\rot
    \delt{1} \bm B]
 \nonumber\\
&-&\frac{\delt{1}\rho}{4\pi \rho^2}\left\{
   [\delt{1}\bm B\times\rot \bm B^{(\mathrm d)}]
   +[\bm B^{(\mathrm d)}\times \rot \delt{1} \bm B]
   \right\}
 \nonumber \\
 &-&\left[\frac{\delt{2}\rho}{4\pi \rho^2}-\frac{(\delt{1}\rho)^2}{4\pi \rho^3}\right][\bm B^{(\mathrm
d)}\times\rot \bm B^{(\mathrm d)}]. \label{MagStress2}
\end{eqnarray}
Here $\delta \bm B=\delt{1} \bm B+\delt{2} \bm B+\ldots=\bm
B^{(\mathrm o)}-\bm B^{(\mathrm d)}$ is Eulerian variation
of magnetic field on course of oscillations and $\bm
B^{(\mathrm o)}$ is total magnetic field (in accurate
solution of MHD equations). However, in contrast to
$B^{(\mathrm d)}$, we are interested only to the first and
second order perturbations of the magnetic field (with
respect of drift solution), assuming Eulerian variation
$\delta B^{(\mathrm o)}$ to be small.
Let us present a trajectory of fluid elements for this
solution in form
\begin{equation}
     \bm c^{(\mathrm o)}(\bm a, t)= \bm c^{(\mathrm d)}(\bm a, t)+
     \bm \xi^{(\mathrm o)}(\bm c^{(\mathrm d)}(\bm a,t),t).
\end{equation}
This form allows us to use the same relations between
Eulerian variation of the velocity $\delt{1} \bm v$ and
$\delt{2} \bm v$ and $\bm \xi^{(\mathrm o)}$ as we do in
absence
of drift motions [Eqs.\ (\ref{delt1v}) and (\ref{delt2v})].%
\footnote{Even the last equalities in Eqs. (\ref{delt1v})
and (\ref{delt2v}) can be applied for unperturbed
stationary configuration corresponding to solid state
rotation as far as drift motion perturb stationary velocity
only in the second order in $\alpha$: $\delt{d} v\sim
\alpha^2 v$.}

 Magnetic field and
trajectory are coupled by the induction equation, which
leads to vanishing Lagrangian perturbation of $\bm
B/\rho$ (see \citealt{ga07}, for example).
Thus, Lagrangian variation $\Delta B^i=-B^i \nabla_k
\xi^{(\mathrm o)\,k}$  and  Eulerian variation of magnetic
field can be derived following general formalism for vector fields (\citealt{fs78a,Sa04}):%
\begin{eqnarray}
 \delt{1}  B^{(\mathrm o)\, i}&=&
  -B^{(\mathrm d)\, i}\nabla_j\ \xi^{(1)\,j}
  -\xi^{(1)\,j}\nabla_j\ B^{(\mathrm d)\, i}
  \nonumber \\
   &+&
   B^{(\mathrm d)\, j}\nabla_j\xi^{(1)\,i},
  \\
 \delt{2}  B^{(\mathrm o)\, i}&=&
 -B^{(\mathrm d)\, i}\nabla_j\ \xi^{(2)\,j}
  +B^{(\mathrm d)\, j}\nabla_j\xi^{(2)\,i}
\nonumber \\
   &-& \xi^{(2)\,j}\nabla_j\ B^{(\mathrm d)\, i}
   -\frac{1}{2}\xi^{(1)\,k}\xi^{(1)\,j}\nabla_j\nabla_k  B^{(\mathrm d)\,i}
  \nonumber \\
   &+&\nabla_k \xi^{(1)\,i}\,\Delta^{(1)} B^{(\mathrm o)\, k}
   -
  \xi^{(1)\,k}\nabla_k \delt 1 B^{i}.
\end{eqnarray}
Equations (\ref{lin_mag1}--\ref{lin_mag3}) and
(\ref{sec_mag1}--\ref{sec_mag3}) differ from linear
equations in nonmagnetized case
[(\ref{lin_nonmag1}--\ref{lin_nonmag3}) and
(\ref{sec_nonmag1}--\ref{sec_nonmag3})] only by magnetic
stresses $\delt{1} \bm F^\mathrm{m}$ and $\delt{2} \bm
F^\mathrm{m}$ in Eqs.\ (\ref{lin_mag1}) and
(\ref{sec_mag1}), respectively. Let us analyse effects of
magnetic field onto nonmagnetized ($B=0$) solutions  by
comparison of magnetic and other stresses in these
equations. Nonmagnetic (hydrodynamical) stresses can be
estimated as
 $\delt{1} F^{\mathrm h}\approx \partial_t \delt{1} v = \mathcal{O}(\alpha) R\omega^2$
for the first order Eq.\ (\ref{lin_mag1}) and as
 $\delt{2} F^{\mathrm h}\approx \partial_t \delt{2} v= \mathcal{O}(\alpha^2) R\omega^2$
for the second order Eq.\ (\ref{sec_mag1}). Before
estimating magnetic stresses let us note, that for $B=0$
solutions with nonvanishing drift, displacement
$(\xi^{\mathrm o})^\phi$ contains terms $\propto \alpha^2
\omega t$, which are finite  [i.e.\
$(\xi^{\mathrm
o})^\phi=\mathcal{O}(1)$] at $t \omega>1/\alpha^2$. For
such solution Eulerian variation $\delt{2} B \sim B
\alpha^2 \omega t $ becomes large ($\delt{2} B \gtrsim B\gg
\delt{1}B$). Presence of large Eulerian variation makes it
doubtful, that the theory based on the expanding of
Eulerian variations
in series in $\alpha$
is appropriate here.%
\footnote{It is worth to note, that large Lagrangian
displacement $\bm \xi^{(\mathrm{o})}$ is not a problem in
nonmagnetized case, because Eulerian variations of all
quantities
are small at any moment of time and thus can be expanded in
series in $\alpha$.}

Thus, let us concentrate on the non-drifting solution at
$B=0$ (see Sec.\ \ref{Sec:nonmag})
and demonstrate that it indeed can be applied to describe
perturbations of the drift solutions (i.e.\ to describe
oscillations at nonstationary background
and $B\ne 0$).
 For this solution Lagrangian
displacement $\bm \xi^{(\mathrm{o})}$ stays always small
$\bm \xi^{(\mathrm{o})}=\mathcal{O}(\alpha)$ and even
$\mathcal{O}(\alpha^2)$ terms do not increase with time,
thus $\delt{1} B=\mathcal{O}(\alpha)\approx \alpha
B^{(\mathrm d)}$ and $\delt{2}
B=\mathcal{O}(\alpha^2)\approx \alpha^2 B^{(\mathrm d)}$.
Now we can estimate magnetic stresses of the first and the
second order [given by Eqs.\ (\ref{MagStress1}) and
(\ref{MagStress2})] as $\delt{1} F^\mathrm{m}\approx \alpha
(B^{(\mathrm d)})^2/(4\pi \rho R)$ and $\delt{2}
F^\mathrm{m}\approx \alpha^2 (B^{(\mathrm d)})^2/(4\pi \rho
R)$ respectively.
Thus, magnetic to nonmagnetic stresses ratio in Eqs.\
(\ref{lin_mag1}) and (\ref{sec_mag1}) is $\delt{1}
F^{\mathrm m}/\delt{1} F^{\mathrm h}\approx\delt{2}
F^{\mathrm m}/\delt{2} F^{\mathrm h}\approx (B^{(\mathrm
d)})^2/(4\pi \rho R^2
\omega^2)=v_\mathrm{A}^2/(R^2\omega^2)$, where
$v_\mathrm{A}=\sqrt{(B^{(\mathrm d)})^2/(4\pi\rho)}$ being
Alfv\'{e}n velocity. As a result, we conclude that magnetic
stresses in both of these equations are negligible, if
$B^{(\mathrm d)}\ll B^\mathrm{crit}\approx R\omega
\sqrt{4\pi \rho}$. For a r-mode in NS and fiducial
parameters ($\rho=4\times 10^{14}$~g\,cm$^{-3}$,
$\omega=4/3\Omega=8\pi/3\nu=5000 (\nu/600\mathrm{
Hz})$~s$^{-1}$, $R=10^6$~cm) we obtain
$B_\mathrm{crit}\approx 2\times 10^{17}$~G.%
\footnote{This estimate slightly differs from one obtained
by \cite{Rezzolla_etal00} based on another physical
arguments.
However, factor of few units difference is not a problem for
order-of-magnitude estimates.}
Thus, $B=0$ non-drifting solution can be used as the first
approximation to describe non-drifting oscillation mode for
magnetized NS [first approximation to the solution of Eqs.\
(\ref{lin_mag1}--\ref{lin_mag3}) and
(\ref{sec_mag1}--\ref{sec_mag3})].%
\footnote{This conclusion cannot be applied for
non-magnetized solutions with nonvanishing drift (mass
transfer velocity $v^{(\mathrm{mt})}\neq 0$, see Sec.\
\ref{Sec:nonmag}).
Indeed, stress governing evolution of $v^{(\mathrm{mt})}$
is zero in nonmagnetized case, but for magnetized star it
should be compared with increasing [because
$v^{(\mathrm{mt})}\neq 0$] magnetic stress $\delt{2}
F^{\mathrm m}$. Thus, magnetic stress can not be ignored
and should modify such solutions.}

Let us analyse corrections to this solution, which can be
associated with magnetic field.
In the first order these effects were studied accurately by
\cite{lee05,arr12,cs13,aly15}. For simplicity, here we
apply analytical estimates, which is enough for the purpose
of the study.
Magnetic stresses, which oscillate along with oscillations,
can modify oscillation frequency slightly by providing
additional stiffness to oscillations.
In analogy with effect of small magnetic field on sound
waves in plasma, we anticipate increasing  of the
oscillation frequency to the value $\delta \omega \approx
\omega\, (v_\mathrm{A}/v_\mathrm{s})^2$, where
$v_\mathrm{s}$ is a phase velocity in absence of magnetic
field. For r-modes, taken $\omega R$ as a phase velocity,
we obtain $\delta \omega/\omega \approx
(B^{(\mathrm d)}/B_\mathrm{crit})^2$.
This estimate is in a reasonable agreement with accurate
calculation by \cite{aly15} and by \cite{cs13} (for
rotation, which is not too
slow).%
\footnote{For very slow rotation ($\nu\lesssim 0.07 \sqrt{G
M/R^3}$) \cite{cs13} suggests $\delta \omega/\omega\propto
B^4$. If this holds true, we can significantly overestimate
the effect of the magnetic field on the oscillation
frequency. However, \cite{aly15} suggest  scaling $\delta
\omega/\omega\propto (B/B_\mathrm{crit})^2$ (in agreement
with our estimates) even for very slow rotation. The
difference between the results of these papers
can be associated with different magnetic field
configuration and models of NSs applied by authors
or, more probably, by too strong magnetic fields $B\gtrsim
10^{15}$~G discussed by \cite{cs13}. For such a large
magnetic field the lowest order expansion $\delta
\omega/\omega\propto B^2$ may become invalid at low
rotation frequencies.}
The terms in $\delt{2} F^{\mathrm m}$ with nonvanishing
oscillation average (for example, $\propto \delt{1}\bm
B\times\rot \delt{1} \bm B$) cannot lead to significant
displacement across the magnetic field, thus significantly
affecting magnetic field, for a simple reason: additional
displacement of second order $\delta
\xi^B=\mathcal{O}(\alpha^2)$ across magnetic field would
produce compensating stress due to deformation of the
magnetic force lines. As a result, we suppose that such
terms do not lead to significant modification of magnetic
field and global drift motion. Concluding, for a given
constant $\bm B^{(\mathrm d)}=\mathrm{const}$, none of
magnetic terms leads to significant modification of
nonmagnetized solution, which would affect evolution of
$B$. In general case $\bm B^{(\mathrm d)}$ is not a
constant, but varies on a timescale
$t^{(\mathrm{d})}\approx R/v^{(\mathrm d)}\approx
1/(\alpha^2\omega)\gg 1/\omega$. Hence, adiabatic invariant
of the first order oscillation equation $E/\omega$ should
be conserved during $\bm B^{(\mathrm d)}$ evolution. As far
as $B^{(\mathrm d)}\ll B_\mathrm{crit}$, corrections to the
oscillation frequency are small and oscillation energy is
also conserved in leading order.

Let us now demonstrate that a general second order solution
can be presented in a form of superposition of drift
solution and non-drifting r-mode. To do that, we analyse
evolution of the oscillation mode with nonvanishing mass
transfer velocity at the initial moment of time.
It is especially important to analyze this possibility as
far as \cite{fll15} clearly demonstrated that r-modes
excited by gravitational radiation in uniformaly rotating
nonmagnetized NS correspond to well defined nonvanishing
mass transfer velocity.
Let us assume that at this moment Eulerian perturbations of
the first order in $\alpha$ were given by an oscillation
mode for initial magnetic field $\bm B^{(0)}$. In this case
an arbitrary perturbations of the second order can be
presented as a sum of second order perturbation,
corresponding to the second order non-drifting solution and
corrections, associated with nonvanishing mass transfer
velocity. Let us use these corrections, which are second
order in $\alpha$, to define initial state and determine
drifting solution $\bm c^{(\mathrm d)}(\bm a, t)$ by
solving exact MHD equations. As long as initial velocity
perturbations are small
[$\delt{d}v=v^{(\mathrm{mt})}=\mathcal{O}(\alpha^2)$], we
expect that temporal evolution of this velocity field takes
place on Alfv\'{e}n timescale, preserving Eulerian
variation of all quantities (with possible exception of the
magnetic field) to be of order of $\alpha^2$, i.e.
trajectory $\bm c^{(\mathrm d)}(\bm a, t)$ satisfies
requirements, imposed
to the drift motion solution in the beginning of section.%
Thus, we present initial condition as a composition of
drift solution and  non-drifting mode. It allows us to
apply this solution to describe evolution of initial
perturbations. Thus secular evolution of the magnetic field
follows the drift solution, i.e. it should be the same as
in absence of non-drifting r-mode. As a result, we can
estimate an upper limit for the enhancement of the magnetic
field energy $\approx\Delta B^2 R^3/6$ as an energy of
initial drift motion $\approx 2\pi R^3\rho (\delt{d} v)
^2/3$. Here $\Delta B^2=B^2-(B^{(0)})^2$. For initial drift
velocity equal to mass transfer velocity $\delt{d}v
=v^{(\mathrm mt)}\approx \alpha^2 R\omega$ and fiducial NS
parameters written above we come to $\sqrt{\Delta B^2}
\approx 10^8 (\alpha/10^{-4})^2$~G. Thus, if initial field
$B\gg 10^8 (\alpha/10^{-4})^2$~G it will not be
enhanced by the drift significantly.%
\footnote{This estimate was suggested by M.E.~Gusakov
during our discussions on the early stage of this work. It
is also in agreement with estimates by \cite{af63} (see
footnote \ref{footnote_af63}).}
If,  on the contrary, initial field $B\ll 10^8
(\alpha/10^{-4})^2$~G, it can be enhanced up to $10^8
(\alpha/10^{-4})^2$~G.

This section's conclusion can be formulated in almost the
same form as for nonmagnetized NS: a general second order
r-mode solution can be presented as a superposition of
oscillating solution with vanishing drift and a solution,
which describes differential rotation in nonoscillating NS.
The main effect of magnetic field is that it modifies
differential rotation in nonoscillating NSs. In case of
magnetized star differential rotation cannot be stationary,
but should evolve on the Alfv\'{e}n timescale.

\section{Discussion and conclusion}
We reanalyze second order effects in r-mode oscillations.
For nonmagnetized NS we refute widely believed statement
that (second order) drift  (secular motion of fluid
elements) is an essential feature of r-modes. We present
the second order r-mode solution with vanishing
drift (non-drifting solution; see Sec.\ \ref{Sec:nonmag}).
Strictly speaking, this solution is applicable in a case of
slow rotation ($\Omega\ll\Omega_\mathrm{K}$) and $\alpha\gg
(\Omega/\Omega_\mathrm{K})^2$. However, we suppose that
these restrictions are not crucial for the main results of
paper  (see footnote \ref{footnote_gen}), but we leave a
strict proof of this conjecture beyond the scope of the
paper.
In Sec.\ \ref{Sec:mag} we demonstrate that this
non-drifting solution
is valid for magnetized NS. On the contrary, $B=0$ r-mode
solutions with nonvanishing drift cannot be applied
directly to magnetized NS because magnetic field leads to
evolution of the mass transfer velocity, which describes
drift, on the Alfv\'{e}n timescale. However this evolution
is decoupled from oscillations. This means that the general
r-mode solution can be presented as superposition of two
solutions: (a) the solution characterizing evolution of
mass transfer velocity (drift motion), which is the same as
evolution of differential rotation in
\textit{nonoscillating} NS (initial conditions correspond
to the mass transfer velocity at $t=0$);  and (b)
non-drifting r-mode solution. Secular evolution of magnetic
field is entirely determined by drift motion, thus its
initial energy gives an upper limit for the enhancement of
magnetic field energy. The total energy of the
r-mode cannot be converted into magnetic energy %
because it exceeds drift energy for a factor of
$1/\alpha^2$.
Thus we conclude that magnetic field energy enhancement
does not prove to be relevant for oscillation energy
decrease and therefore cannot prevent instability.

Still, in our consideration we neglect gravitational
radiation-reaction force. As shown by \cite{fll15}
(published in arXiv, while the present manuscript has been
under revision), this force excites r-mode with well
defined nonvanishing mass transfer velocity (differential
rotation) in nonmagnetized NSs. Strictly speaking, in the
present paper we demonstrate that this mode does not dump
regardless of magnetic field  enhancement (the opposite to
what follows from results by
\cite{Rezzolla_etal00,Rezzolla_etal01a,Rezzolla_etal01b},
in case when gravitational radiation force stop acting. In
reality, gravitational radiation force acts permanently and
accurate description of second order r-modes under combined
action of gravitational radiation and magnetic field is
particularly interesting task. We expect that two limiting cases are possible,
depending on the ratio of Alfv\'{e}n timescale $t_\mathrm{A}$ to
r-mode growing time $\tau$.
If $t_\mathrm{A}\ll \tau$, then magnetic field has enough time
to affect mass transfer rate and the theory described here remains applicable.
In the opposite case, namely when $t_\mathrm{A}\gg
\tau$, r-modes will saturate rapidly, since they are almost unaffected
by the magnetic field.

Saturated r-modes require special consideration.
It is typically assumed that saturation of r-mode
instability is associated with the lowest parametric
instability threshold in nonlineary coupled triplets of
oscillation modes
[\citealt*{schenk_et_al_02,afmstw03,btw04a,btw04b,btw07,btw09,bw13}].
Within this model, saturated r-mode is always accompanied
by excited daughter modes (at least two).
If amplitudes  of these latter modes are comparable with r-mode
amplitude,
the first order velocity field should be presented as a sum
of three oscillation modes, and each of them affects second
order equations.
Then the total mass transfer velocity, in principle, can
differ from Eq.\ (\ref{vd_Sa}), so that the very existence of
nondrifting solution might be challenged as well as
applicability of the theory discussed here.
However, if saturation amplitude is not too large
($\alpha\ll 1$), we expect that such scenario does not occur and we can obtain
nondrifting solution even for arbitrary number
of eigenmodes excited in first order. However, a more
detailed analysis of this question is left beyond the scope of this
paper.

Strictly speaking, the present results are accurate only at
second order in mode amplitude. However, the drift of the
fluid elements in the next orders (the third and higher)
cannot be excluded. Even if mass transfer velocity for such
drift is very small $v^{(\mathrm mt)}_3\approx
\alpha^3\Omega R\approx 4\times 10^{-3} (\nu/600\mathrm{\,
Hz})\,(\alpha/10^{-4})^3$~cm\,s$^{-1}$, it can lead to
large Lagrangian displacements on a timescale of years (in
nonmagnetic case). The second order drift and first order
oscillations are coupled in the third order. As a result
the theory in third order can be much more complicated than
the one in second order described here.
Furthermore, for very large amplitudes $\alpha\gtrsim 3$,
which can be studied only numerically, r-modes  decay
nonlinearly and strong differential rotation develops
(\citealt{Gressman_etal02,Kastaun11}). Such a strong
differential rotation (angular velocities in the range
$0.5\ldots 1.2$ of the initial one according to
\citealt{Kastaun11}) can strongly affect magnetic field
(formally, it does not contradict with our estimates, which
predict enhancement of the magnetic field up to $\approx
10^{17}$~G for $\alpha\approx 3$).
However, r-mode saturation amplitudes calculated by
\cite{btw09,bw13} are small $\alpha\ll 1$. Thus it seems
reasonable to restrict consideration to small amplitudes.
In this case, the total mass transfer velocity has small
value of the second order in $\alpha$, even if we include
high order terms. As a result, a general argument by
\cite{af63} (see footnote \ref{footnote_af63}), that too
slow differential rotation cannot generate magnetic field,
is still valid within high order perturbation theory.
Consequently, high order corrections can hardly result in
strong magnetic field generation. However, we leave a more
accurate analysis of high order effects outside the paper.

To conclude we note, that results discussed here can be
easily generalized for any oscillating mode in any
magnetized medium, if secular drift is a perturbation,
allowed by the linear perturbation theory. This property
seems to be rather general, at least it holds true for
r-modes and ocean waves (\citealt{LH53,Moore70}) and we
plan to study whether it is indeed a general property of
second order oscillations or not.

\section*{Acknowledgements}
I'm  grateful to M.~E.~Gusakov and D.~P.~Barsukov for
fruitful discussions and critical remarks, to an anonymous
referee for useful comments, which helped a lot in refining
the paper, and to O.~V.~Zakutnyaya for assistance in
preparation of the manuscript. This work was supported by
RSCF grant 14-12-00316.

\bibliographystyle{mn2e}
\bibliographystyle{mn2e}

\begin{thebibliography}{}

\bibitem[\protect\citeauthoryear{{Abbassi}, {Rieutord} \& {Rezania}}{{Abbassi}
  et~al.}{2012}]{arr12}
{Abbassi} S.,  {Rieutord} M.,    {Rezania} V.,  2012,
\mnras, 419, 2893

\bibitem[\protect\citeauthoryear{Alfv\'{e}n \& Felthammar}{Alfv\'{e}n \&
  Felthammar}{1963}]{af63}
Alfv\'{e}n H.,  Felthammar G.,  1963, Cosmical
electrodynamics. Clarendon, Oxford

\bibitem[\protect\citeauthoryear{{Andersson}}{{Andersson}}{1998}]{andersson98}
{Andersson} N.,  1998, \apj, 502, 708

\bibitem[\protect\citeauthoryear{{Andersson} \& {Kokkotas}}{{Andersson} \&
  {Kokkotas}}{2001}]{ak01}
{Andersson} N.,  {Kokkotas} K.~D.,  2001, International
Journal of Modern
  Physics D, 10, 381

\bibitem[\protect\citeauthoryear{{Andersson}, {Kokkotas} \&
  {Stergioulas}}{{Andersson} et~al.}{1999}]{aks99}
{Andersson} N.,  {Kokkotas} K.~D.,    {Stergioulas} N.,
1999, \apj, 516, 307

\bibitem[\protect\citeauthoryear{{Arras}, {Flanagan}, {Morsink}, {Schenk},
  {Teukolsky} \& {Wasserman}}{{Arras} et~al.}{2003}]{afmstw03}
{Arras} P.,  {Flanagan} E.~E.,  {Morsink} S.~M.,  {Schenk}
A.~K.,  {Teukolsky}
  S.~A.,    {Wasserman} I.,  2003, \apj, 591, 1129

\bibitem[\protect\citeauthoryear{{Asai}, {Lee} \& {Yoshida}}{{Asai}
  et~al.}{2015}]{aly15}
{Asai} H.,  {Lee} U.,    {Yoshida} S.,  2015, ArXiv
e-prints

\bibitem[\protect\citeauthoryear{{Bildsten}}{{Bildsten}}{1998}]{bildsten98}
{Bildsten} L.,  1998, \apjl, 501, L89

\bibitem[\protect\citeauthoryear{{Bondarescu}, {Teukolsky} \&
  {Wasserman}}{{Bondarescu} et~al.}{2007}]{btw07}
{Bondarescu} R.,  {Teukolsky} S.~A.,    {Wasserman} I.,
2007, \prd, 76, 064019

\bibitem[\protect\citeauthoryear{{Bondarescu}, {Teukolsky} \&
  {Wasserman}}{{Bondarescu} et~al.}{2009}]{btw09}
{Bondarescu} R.,  {Teukolsky} S.~A.,    {Wasserman} I.,
2009, \prd, 79, 104003

\bibitem[\protect\citeauthoryear{{Bondarescu} \& {Wasserman}}{{Bondarescu} \&
  {Wasserman}}{2013}]{bw13}
{Bondarescu} R.,  {Wasserman} I.,  2013, \apj, 778, 9

\bibitem[\protect\citeauthoryear{{Brink}, {Teukolsky} \& {Wasserman}}{{Brink}
  et~al.}{2004a}]{btw04a}
{Brink} J.,  {Teukolsky} S.~A.,    {Wasserman} I.,  2004a,
\prd, 70, 124017

\bibitem[\protect\citeauthoryear{{Brink}, {Teukolsky} \& {Wasserman}}{{Brink}
  et~al.}{2004b}]{btw04b}
{Brink} J.,  {Teukolsky} S.~A.,    {Wasserman} I.,  2004b,
\prd, 70, 121501

\bibitem[\protect\citeauthoryear{{Chandrasekhar}}{{Chandrasekhar}}{1970}]{chandrasekhar70a}
{Chandrasekhar} S.,  1970, \prl, 24, 611

\bibitem[\protect\citeauthoryear{{Chirenti} \& {Sk{\'a}kala}}{{Chirenti} \&
  {Sk{\'a}kala}}{2013}]{cs13}
{Chirenti} C.,  {Sk{\'a}kala} J.,  2013, \prd, 88, 104018

\bibitem[\protect\citeauthoryear{Chirenti, Sk\'akala \& Yoshida}{Chirenti
  et~al.}{2013}]{csy13}
Chirenti C.,  Sk\'akala J.,    Yoshida S.,  2013, Phys.
Rev. D, 87, 044043

\bibitem[\protect\citeauthoryear{{Chirenti}, {Sk{\'a}kala} \&
  {Yoshida}}{{Chirenti} et~al.}{2014}]{csy14}
{Chirenti} C.,  {Sk{\'a}kala} J.,    {Yoshida} S.,  2014,
Astronomische
  Nachrichten, 335, 618

\bibitem[\protect\citeauthoryear{{Chugunov}, {Gusakov} \& {Kantor}}{{Chugunov}
  et~al.}{2014}]{cgk14}
{Chugunov} A.~I.,  {Gusakov} M.~E.,    {Kantor} E.~M.,
2014, \mnras, 445, 385

\bibitem[\protect\citeauthoryear{{Friedman}, {Lindblom} \&
  {Lockitch}}{{Friedman} et~al.}{2015}]{fll15}
{Friedman} J.~L.,  {Lindblom} L.,    {Lockitch} K.~H.,
2015, ArXiv e-prints

\bibitem[\protect\citeauthoryear{{Friedman} \& {Morsink}}{{Friedman} \&
  {Morsink}}{1998}]{fm98}
{Friedman} J.~L.,  {Morsink} S.~M.,  1998, \apj, 502, 714

\bibitem[\protect\citeauthoryear{{Friedman} \& {Schutz}}{{Friedman} \&
  {Schutz}}{1978a}]{fs78a}
{Friedman} J.~L.,  {Schutz} B.~F.,  1978a, \apj, 221, 937

\bibitem[\protect\citeauthoryear{{Friedman} \& {Schutz}}{{Friedman} \&
  {Schutz}}{1978b}]{fs78b}
{Friedman} J.~L.,  {Schutz} B.~F.,  1978b, \apj, 222, 281

\bibitem[\protect\citeauthoryear{{Glampedakis} \& {Andersson}}{{Glampedakis} \&
  {Andersson}}{2007}]{ga07}
{Glampedakis} K.,  {Andersson} N.,  2007, \mnras, 377, 630

\bibitem[\protect\citeauthoryear{{Gressman}, {Lin}, {Suen}, {Stergioulas} \&
  {Friedman}}{{Gressman} et~al.}{2002}]{Gressman_etal02}
{Gressman} P.,  {Lin} L.-M.,  {Suen} W.-M.,  {Stergioulas}
N.,    {Friedman}
  J.~L.,  2002, \prd, 66, 041303

\bibitem[\protect\citeauthoryear{{Gusakov}, {Chugunov} \& {Kantor}}{{Gusakov}
  et~al.}{2014a}]{gck14b}
{Gusakov} M.~E.,  {Chugunov} A.~I.,    {Kantor} E.~M.,
2014a, \prd, 90, 063001

\bibitem[\protect\citeauthoryear{{Gusakov}, {Chugunov} \& {Kantor}}{{Gusakov}
  et~al.}{2014b}]{gck14a}
{Gusakov} M.~E.,  {Chugunov} A.~I.,    {Kantor} E.~M.,
2014b, \prl, 112,
  151101

\bibitem[\protect\citeauthoryear{{Haskell}, {Degenaar} \& {Ho}}{{Haskell}
  et~al.}{2012}]{hdh12}
{Haskell} B.,  {Degenaar} N.,    {Ho} W.~C.~G.,  2012,
\mnras, 424, 93

\bibitem[\protect\citeauthoryear{{Hessels}, {Ransom}, {Stairs}, {Freire},
  {Kaspi} \& {Camilo}}{{Hessels} et~al.}{2006}]{hrsfkc06}
{Hessels} J.~W.~T.,  {Ransom} S.~M.,  {Stairs} I.~H.,
{Freire} P.~C.~C.,
  {Kaspi} V.~M.,    {Camilo} F.,  2006, Science, 311, 1901

\bibitem[\protect\citeauthoryear{{Ho}, {Andersson} \& {Haskell}}{{Ho}
  et~al.}{2011}]{hah11}
{Ho} W.~C.~G.,  {Andersson} N.,    {Haskell} B.,  2011,
\prl, 107, 101101

\bibitem[\protect\citeauthoryear{{Karino}, {Yoshida} \& {Eriguchi}}{{Karino}
  et~al.}{2001}]{kye01}
{Karino} S.,  {Yoshida} S.,    {Eriguchi} Y.,  2001, \prd,
64, 024003

\bibitem[\protect\citeauthoryear{{Kastaun}}{{Kastaun}}{2011}]{Kastaun11}
{Kastaun} W.,  2011, \prd, 84, 124036


\bibitem[\protect\citeauthoryear{{Kinney} \& {Mendell}}{{Kinney} \&
  {Mendell}}{2003}]{km03}
{Kinney} J.~B.,  {Mendell} G.,  2003, \prd, 67, 024032

\bibitem[\protect\citeauthoryear{{Lee}}{{Lee}}{2005}]{lee05}
{Lee} U.,  2005, \mnras, 357, 97

\bibitem[\protect\citeauthoryear{{Lindblom}, {Owen} \& {Morsink}}{{Lindblom}
  et~al.}{1998}]{lom98}
{Lindblom} L.,  {Owen} B.~J.,    {Morsink} S.~M.,  1998,
\prl, 80, 4843

\bibitem[\protect\citeauthoryear{{Lindblom}, {Tohline} \&
  {Vallisneri}}{{Lindblom} et~al.}{2001}]{LTV01}
{Lindblom} L.,  {Tohline} J.~E.,    {Vallisneri} M.,  2001,
Physical Review
  Letters, 86, 1152

\bibitem[\protect\citeauthoryear{Lockitch, Andersson \& Friedman}{Lockitch
  et~al.}{2000}]{laf00}
Lockitch K.~H.,  Andersson N.,    Friedman J.~L.,  2000,
Phys. Rev. D, 63,
  024019

\bibitem[\protect\citeauthoryear{{Lockitch}, {Friedman} \&
  {Andersson}}{{Lockitch} et~al.}{2003}]{lfa03}
{Lockitch} K.~H.,  {Friedman} J.~L.,    {Andersson} N.,
2003, \prd, 68, 124010

\bibitem[\protect\citeauthoryear{{Longuet-Higgins}}{{Longuet-Higgins}}{1953}]{LH53}
{Longuet-Higgins} M.~S.,  1953, Phil. Trans. R. Soc. Long.
A, 245, 535

\bibitem[\protect\citeauthoryear{{Mendell}}{{Mendell}}{2001}]{mendell01}
{Mendell} G.,  2001, \prd, 64, 044009

\bibitem[\protect\citeauthoryear{{Moore}}{{Moore}}{1970}]{Moore70}
{Moore} D.,  1970, Geophysical and Astrophysical Fluid
Dynamics, 1, 237

\bibitem[\protect\citeauthoryear{{Owen}, {Lindblom}, {Cutler}, {Schutz},
  {Vecchio} \& {Andersson}}{{Owen} et~al.}{1998}]{olcsva98}
{Owen} B.~J.,  {Lindblom} L.,  {Cutler} C.,  {Schutz}
B.~F.,  {Vecchio} A.,
  {Andersson} N.,  1998, \prd, 58, 084020

\bibitem[\protect\citeauthoryear{{Provost}, {Berthomieu} \& {Rocca}}{{Provost}
  et~al.}{1981}]{pbr81}
{Provost} J.,  {Berthomieu} G.,    {Rocca} A.,  1981, \aap,
94, 126

\bibitem[\protect\citeauthoryear{{Rezzolla}, {Lamb}, {Markovi{\'c}} \&
  {Shapiro}}{{Rezzolla} et~al.}{2001a}]{Rezzolla_etal01a}
{Rezzolla} L.,  {Lamb} F.~K.,  {Markovi{\'c}} D., {Shapiro}
S.~L.,  2001a,
  \prd, 64, 104013

\bibitem[\protect\citeauthoryear{{Rezzolla}, {Lamb}, {Markovi{\'c}} \&
  {Shapiro}}{{Rezzolla} et~al.}{2001b}]{Rezzolla_etal01b}
{Rezzolla} L.,  {Lamb} F.~K.,  {Markovi{\'c}} D., {Shapiro}
S.~L.,  2001b,
  \prd, 64, 104014

\bibitem[\protect\citeauthoryear{{Rezzolla}, {Lamb} \& {Shapiro}}{{Rezzolla}
  et~al.}{2000}]{Rezzolla_etal00}
{Rezzolla} L.,  {Lamb} F.~K.,    {Shapiro} S.~L.,  2000,
\apjl, 531, L139

\bibitem[\protect\citeauthoryear{Ruoff \& Kokkotas}{Ruoff \&
  Kokkotas}{2002}]{rk02}
Ruoff J.,  Kokkotas K.~D.,  2002, Monthly Notices of the
Royal Astronomical
  Society, 330, 1027

\bibitem[\protect\citeauthoryear{{S{\'a}}}{{S{\'a}}}{2004}]{Sa04}
{S{\'a}} P.~M.,  2004, \prd, 69, 084001

\bibitem[\protect\citeauthoryear{{S{\'a}} \& {Tom{\'e}}}{{S{\'a}} \&
  {Tom{\'e}}}{2005}]{ST05}
{S{\'a}} P.~M.,  {Tom{\'e}} B.,  2005, \prd, 71, 044007

\bibitem[\protect\citeauthoryear{{Schenk}, {Arras}, {Flanagan}, {Teukolsky} \&
  {Wasserman}}{{Schenk} et~al.}{2002}]{schenk_et_al_02}
{Schenk} A.~K.,  {Arras} P.,  {Flanagan} {\'E}.~{\'E}.,
{Teukolsky} S.~A.,
  {Wasserman} I.,  2002, \prd, 65, 024001

\bibitem[\protect\citeauthoryear{{Spruit}}{{Spruit}}{1999}]{Spruit99_DifRot}
{Spruit} H.~C.,  1999, \aap, 349, 189

\bibitem[\protect\citeauthoryear{{Stergioulas} \& {Font}}{{Stergioulas} \&
  {Font}}{2001}]{SF01}
{Stergioulas} N.,  {Font} J.~A.,  2001, Physical Review
Letters, 86, 1148

\bibitem[\protect\citeauthoryear{{Yakovlev} \& {Shalybkov}}{{Yakovlev} \&
  {Shalybkov}}{1991}]{ys91}
{Yakovlev} D.~G.,  {Shalybkov} D.~A.,  1991, \apss, 176,
191

\end{thebibliography}

\label{lastpage}

\end{document}